\documentclass[aps,prd,twocolumn,preprintnumbers,superscriptaddress,nofootinbib,floatfix]{revtex4-1}

\usepackage{amsmath,amssymb}
\usepackage{bm}
\usepackage{graphicx}
\usepackage{epstopdf}
\usepackage{hyperref}
\usepackage{array}
\usepackage[utf8]{inputenc}
\usepackage{soul}
\usepackage[usenames, dvipsnames]{color}

\usepackage{subfigure}
\usepackage{slashed}
\usepackage{afterpage}
\usepackage{psfrag}
\usepackage{framed}
\usepackage{multirow}

\enlargethispage{\baselineskip}

\hypersetup{
     colorlinks   = true,
     citecolor    = blue
}

\newcommand{\be}{\begin{equation}}
\newcommand{\ee}{\end{equation}}
\newcommand{\bea}{\begin{eqnarray}}
\newcommand{\eea}{\end{eqnarray}}
\newcommand{\bal}{\begin{align} }
\newcommand{\eal}{ \end{align} }

\newcommand{\exclude}[1]{}

\begin{document}

\title{Dark Matter implications of DAMA/LIBRA-phase2 results}

\newcommand{\FIRSTAFF}{\affiliation{The Oskar Klein Centre for Cosmoparticle Physics,
	Department of Physics,\\
	Stockholm University,
	AlbaNova,
	10691 Stockholm,
	Sweden}}
\newcommand{\SECONDAFF}{\affiliation{Nordita,
	KTH Royal Institute of Technology and Stockholm University,
	Roslagstullsbacken 23,
	10691 Stockholm,
	Sweden}}
\newcommand{\THIRDAFF}{\affiliation{Leinweber Center for Theoretical Physics, 
	Department of Physics,
	University of Michigan,
	Ann Arbor,
	MI 48109,
	USA}}
\newcommand{\FOURTHAFF}{\affiliation{Department of Physics,
	University of North Florida,
	Jacksonville, FL 32224, USA}}

\author{Sebastian Baum}
\email[Electronic address: ]{sbaum@fysik.su.se}
\FIRSTAFF
\SECONDAFF

\author{Katherine Freese}
\email[Electronic address: ]{ktfreese@umich.edu}
\FIRSTAFF
\SECONDAFF
\THIRDAFF

\author{Chris Kelso}
\email[Electronic address: ]{ckelso@unf.edu}
\FOURTHAFF

\preprint{LCTP-18-08}
\preprint{NORDITA-2018-026}

\date{\today}

\begin{abstract}
Recently, the DAMA/LIBRA collaboration released updated results from their search for the annual modulation signal from Dark Matter (DM) scattering in the detector. Besides approximately doubling the exposure of the DAMA/LIBRA data set, the updated photomultiplier tubes of the experiment allow a lower recoil energy threshold of 1\,keV electron equivalent compared to the previous threshold of 2 keV electron equivalent. We study the compatibility of the observed modulation signal with DM scattering. Due to a conspiracy of multiple effects, the new data at low recoil energies is very powerful for testing the DM hypothesis. We find that canonical (isospin conserving) spin-independent DM-nucleon interactions are no longer a good fit to the observed modulation signal in the standard halo model. The canonical spin-independent case is disfavored by the new data, with best fit points of a DM mass of $\sim 8\,$GeV, disfavored by $5.2\,\sigma$, or a mass of $\sim 54\,$GeV, disfavored by $2.5\,\sigma$. Allowing for isospin violating spin independent interactions, we find a region with a good fit to the data with suppressed effective couplings to iodine for DM masses of $\sim 10\,$GeV. We also consider spin-dependent DM-nucleon interactions, which yield good fits for similar DM masses of $\sim 10\,$GeV or $\sim 45\,$GeV.
\end{abstract}

\maketitle

\section{Introduction}
From the first half of the 20th century until today, evidence of the gravitational effects of Dark Matter (DM) has been collected at length scales ranging from the smallest known galaxies to the largest observable scales of our Universe. However, despite significant experimental efforts, the nature of DM is unknown as yet, and no conclusive evidence for interactions of DM with ordinary matter beyond gravity has been observed. In particular, the null results from direct detection experiments searching for the scattering of Weakly Interacting Massive Particle (WIMP) DM off target nuclei in the detector have set stringent bounds on the interaction strength~\cite{Angloher:2015ewa,Amole:2017dex,Akerib:2017kat,Aprile:2017iyp,Agnese:2017jvy,Cui:2017nnn}. 

However, there is one long-standing exception to the null results from direct detection experiments: the DAMA/LIBRA collaboration has been reporting an excess in their DM search for more than a decade, claiming a statistical significance of more than $9\,\sigma$~\cite{Bernabei:2008yi,Bernabei:2010mq,Bernabei:2013xsa}, compatible with a $\sim 10\,$GeV or $\sim 70\,$GeV WIMP~\cite{Gondolo:2005hh,Petriello:2008jj,Chang:2008xa,Fairbairn:2008gz,Savage:2008er,Bottino:2008mf}. The approach of the DAMA/LIBRA collaboration is radically different from those of most other WIMP direct detection experiments, which attempt to measure the total rate of nuclear recoils induced by WIMPs scattering off the target nuclei. Measuring the total rate requires extremely good rejection of background events as well as knowledge of the remaining backgrounds. The DAMA/LIBRA collaboration on the other hand attempts to measure the annual modulation of the WIMP-nucleus recoil rate due to the motion of the Earth around the Sun~\cite{Drukier:1986tm,Freese:2012xd}. Although the rate of the modulated signal is much smaller than the total scattering rate, this approach is advantageous because it offers powerful background rejection: most backgrounds are not expected to be modulated, and if any backgrounds would be modulated their period should differ from 1\,yr and/or their phase should be different of that of the DM signal for which the maximum for most recoil energies is expected around June 2nd. 

In any given WIMP model the overall and the modulated direct detection rate are directly related. Albeit comparing the null results from the remaining direct detection experiments with the detection claim from the DAMA/LIBRA collaboration is a challenging statistics problem, there is strong tension between the DAMA/LIBRA claim and the null results for the canonical spin-independent and spin-dependent WIMP-nucleus interactions, see e.g. Refs.~\cite{Savage:2008er,Savage:2009mk,Savage:2010tg}. In light of this tension, many attempts have been made to reconcile the DAMA/LIBRA claim with the null results using other types of interactions, e.g. all allowed operators in the non-relativistic effective field theory for WIMP-nucleus scattering~\cite{Fitzpatrick:2012ix,Catena:2016hoj}, or by using halo-independent methods~\cite{McCabe:2011sr}. In addition, numerous proposals to explain the signal observed by DAMA/LIBRA by modulated backgrounds have been brought forward, all of which have been refuted by the DAMA/LIBRA collaboration, see e.g.~\cite{Pradler:2012qt,Pradler:2012bf,Pradler:2012bf,Davis:2014cja,Bernabei:2014tqa,Klinger:2015vga}.

Recently, the DAMA/LIBRA collaboration has released first results from the upgraded {\it DAMA/LIBRA-phase2} experiment~\cite{DAMAp2, Bernabei:2018yyw}. Besides almost doubling the total exposure of the DAMA/LIBRA experiment, the upgraded detector allows for a lower recoil energy threshold of 1\,keV electron equivalent (keV$_{\rm ee}$) compared to the 2\,keV$_{\rm ee}$ threshold of DAMA/LIBRA-phase1. 

Extending the reach of the DAMA/LIBRA detector to lower recoil energies below $\sim 2\,$keV$_{\rm ee}$  is crucial for testing if the observed signal is indeed compatible with DM, as discussed in Refs.~\cite{Chang:2008xa,Kelso:2013gda}.  Due to a conspiracy of multiple effects, the behavior of the signals for both the $\sim 10$\,GeV and the $\sim 70\,$GeV mass hypotheses radically changes below $\sim 2\,$keV$_{\rm ee}$ in the standard halo model. For the low mass hypothesis, the modulation signal above $~\sim 2\,$keV$_{\rm ee}$ is predominantly due to scattering off sodium nuclei in DAMA/LIBRA's NaI target. However, for smaller recoil energies the recoil amplitude for scattering off sodium nuclei decreases, approaching the phase reversal~\cite{Primack:1988zm,Hasenbalg:1998zf,Lewin:1995rx,Green:2003yh,Lewis:2003bv} described further below in Sec.~\ref{sec:AnnMod}. Simultaneously the overall scattering rate as well as the modulation signal from scattering off iodine target nuclei increases rapidly, leading to an expected sharp rise of the total observed modulation amplitude below $\sim 2\,$keV$_{\rm ee}$. On the other hand, for the high mass hypothesis a decrease of the recoil rate is expected below $\sim 2.5\,$keV$_{\rm ee}$: the modulation signal for $\sim 70\,$GeV WIMPs is predominately due to scattering off iodine target nuclei for the energy range observed by DAMA/LIBRA and decreases below $\sim 2.5\,$keV$_{\rm ee}$, approaching phase reversal at recoil energies of $\sim 1.5\,$keV$_{\rm ee}$.

In this paper we consider three possible types of DM interactions with the DAMA detector: canonical spin-independent (SI), SI isospin violating (IV), and spin-dependent (SD) interactions, all within the standard halo model. Although in all cases the DAMA/LIBRA modulation results appear to be in severe tension with the null results from other direct detection searches, we demonstrate in this work what can be learned about the DM interpretation of the observed modulation signal from the lowered recoil energy threshold.

Regarding our results:  we will show that in light of the new DAMA/LIBRA-phase2 data for recoil energies below 2\,keV$_{\rm ee}$ together with the changes of the measured modulation amplitude with respect to the DAMA/LIBRA-phase1 data, canonical (isospin conserving) SI WIMP-nucleus interactions are no longer a good fit to the observed modulation signal.  The SI case is disfavored by $5.2\,\sigma$ for the low mass hypothesis and $2.5\,\sigma$ for the high mass hypothesis. Allowing for IV couplings we find a good fit to data for DM masses of $\sim10\,$GeV.  As we will show, the key to improving the fit by using IV couplings is that  the effective coupling to iodine nuclei is suppressed. For SD scattering we consider both proton only or neutron only couplings as well as the mixed case. In all SD cases, we find a low mass region around $\sim 10\,$GeV and a high mass region around $\sim 45\,$GeV, at least one of which yields a good fit to the DAMA/LIBRA-phase2 data for each of the SD cases considered.

The remainder of this paper is organized as follows: In Sec.~\ref{sec:AnnMod} we briefly review the annual modulation signature with special emphasis on the phase reversal. In Sec.~\ref{sec:Analysis} we review our analysis strategy for the compatibility of the modulation signal observed at DAMA/LIBRA being due to DM. In Sec.~\ref{sec:Results} we present the results from this analysis. We reserve Sec.~\ref{sec:Conclusions} for our conclusions.

\section{Annual Modulation}\label{sec:AnnMod}

\begin{figure*}
	\includegraphics[width=0.49\linewidth]{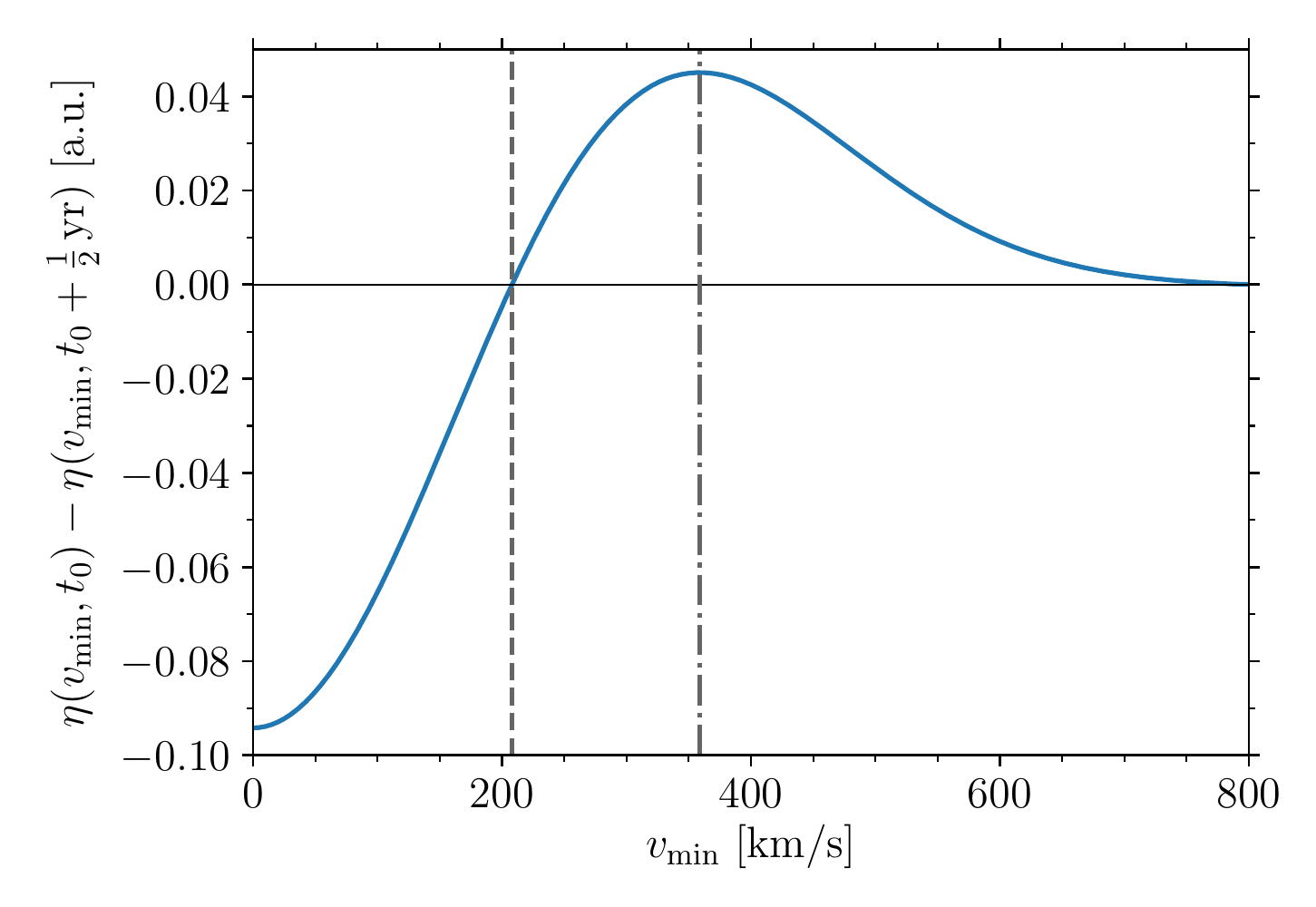}
	\includegraphics[width=0.49\linewidth]{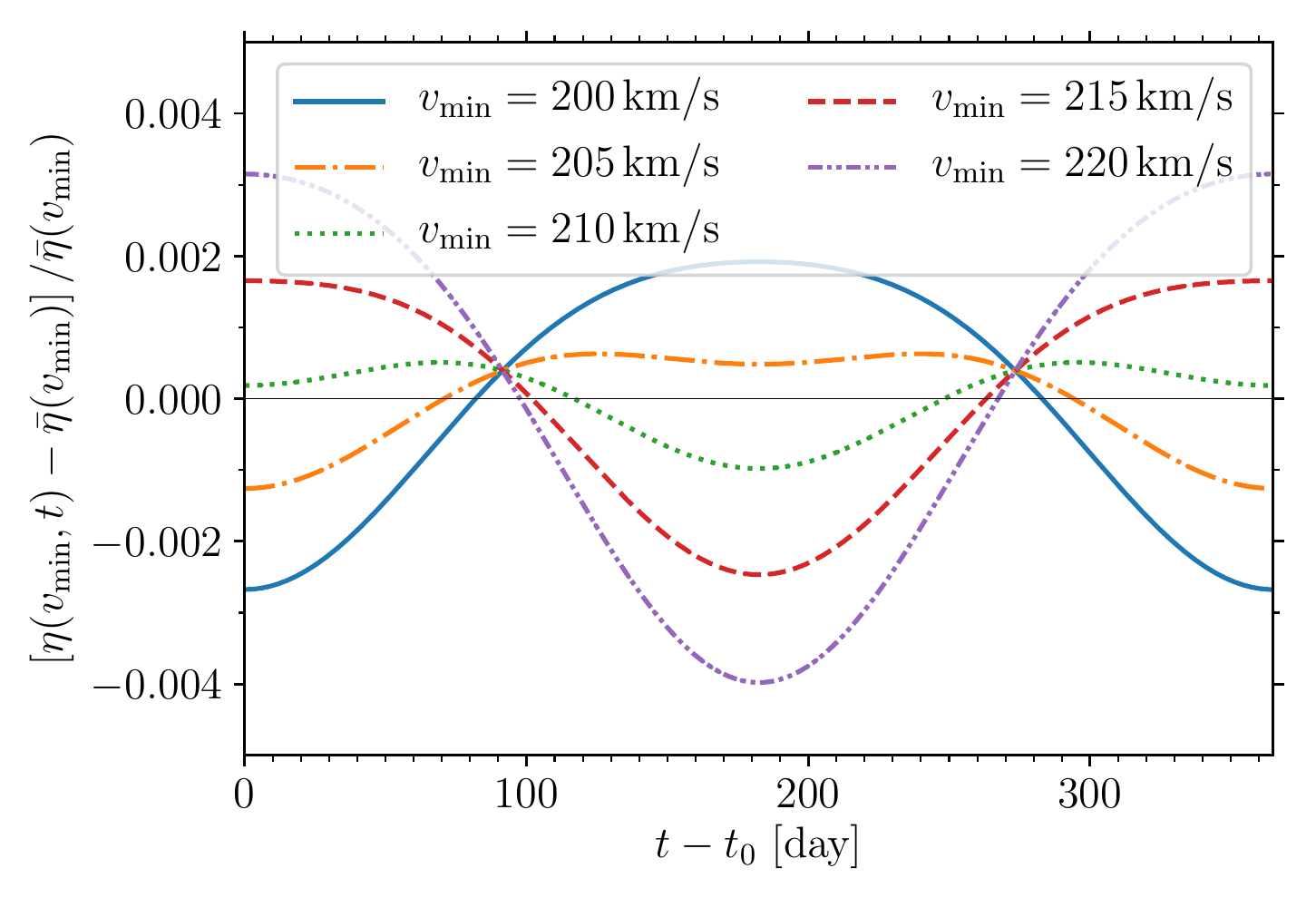}
	\caption{{\it Left:} Amplitude of the modulation (in arbitrary units) as a function of $v_{\rm min}$. The vertical gray dashed and dash dotted lines indicated the values of $v_{\rm min}$ for which the phase flip occurs $v_{\rm min} \approx 210\,$km/s and for which the modulation amplitude becomes maximal $v_{\rm min} \approx 360\,$km/s. {\it Right:} Time dependence of the modulation signal parametrized as the relative difference of the mean inverse speed $\eta$ defined in Eq.~\eqref{eq:eta} from its average value as a function of time over a year for different values of $v_{\rm min}$ indicated in the legend.} 
	\label{fig:mod}
\end{figure*}

The differential nuclear recoil rate per unit detector mass for elastic scattering of WIMPs of mass $m_\chi$ of target nuclei with mass $M$ is 
\begin{equation} \label{eq:diffRate}
	\frac{dR}{dE_{\rm nr}} = 2\frac{\rho_\chi}{m_\chi} \int d^3v\,v f(\mathbf{v},t) \frac{d\sigma}{dq^2}(q^2,v) \;,
\end{equation}
where $\rho_\chi$ is the local WIMP density, $f(\mathbf{v},t)$ the time-dependent WIMP velocity distribution, and $d\sigma/dq^2(q^2,v)$ the differential DM-nucleus scattering cross section with the momentum exchange $q^2 = 2 M E_{\rm nr}$. The nuclear recoil energy $E_{\rm nr} = \left(\mu^2 v^2/M\right)\left(1-\cos\theta\right)$ is related to speed of the WIMP relative to the nucleus $v$ and the scattering angle in the center-of-mass frame $\theta$, where $\mu \equiv m_\chi M / (m_\chi+M)$ is the reduced mass of the WIMP-nucleus system. 

For detectors with multiple target materials $T$ the differential recoil rate per unit detector mass is given by
\begin{equation}
	\frac{dR}{d E_{\rm nr}} = \sum\limits_{T} \xi_T \left( \frac{dR}{dE_{\rm nr}} \right)_T \;,
\end{equation}
where $\xi_T$ is the mass fraction of the target material $T$ in the detector and the differential recoil rate for the target $T$ is computed from Eq.~\eqref{eq:diffRate}.

For the canonical spin-independent (SI) and spin-dependent (SD) WIMP-nucleus interactions arising in most particle physics models from the WIMP-quark couplings, the differential cross section has the form
\begin{equation} \label{eq:diffXsec}
	\frac{d\sigma}{dq^2}(q^2,v) = \frac{\sigma_0}{4\mu^2 v^2}F^2(q) \Theta(q_{\rm max}-q) \;,
\end{equation}
where $\sigma_0$ is the WIMP-nucleus scattering cross section at zero momentum transfer. The nuclear form factor $F(q)$ accounts for the finite size of the nucleus, see e.g.~\cite{Helm:1956zz,Lewin:1995rx,Duda:2006uk} for the form factor in the SI case and~\cite{Bednyakov:2004xq,Bednyakov:2006ux} for the SD case. The Heaviside step function $\Theta(q_{\rm max} - q)$ with the maximal momentum transfer $q_{\rm max} = 2\mu v$ can be traded for a lower cutoff $v_{\rm min} = \sqrt{M E_{\rm nr}/2\mu^2}$ in the velocity integral in Eq.~\eqref{eq:diffRate}. Note, that Eq.~\eqref{eq:diffXsec} allows to rewrite the differential recoil rate as
\begin{equation} \label{eq:diffRate2}
	\frac{dR}{dE_{\rm nr}} = \frac{\sigma_0 \rho_\chi}{2\mu^2 m_\chi}F^2(q) \int\limits_{v_{\rm min}} d^3v\,\frac{f(\mathbf{v},t)}{v} \;.
\end{equation}

For SI scattering, the zero momentum WIMP-nucleus cross section can be parameterized as
\begin{equation}
	\sigma_0^{\rm SI} = \frac{4}{\pi} \mu^2 \left[ Z f_p + \left( A-Z \right) f_n \right]^2 \;,
\end{equation}
where $A$ ($Z$) is the number of nucleons (protons) in the nucleus, and $f_p$ ($f_n$) are the effective WIMP-proton (WIMP-neutron) couplings. In many WIMP models one finds $f_p \simeq f_n$, leading to the typical $A^2$ enhancement of the WIMP-nucleus scattering cross section. However, isospin-violating interactions where $f_p$ and $f_n$ differ substantially can alter the expected signal~\cite{Feng:2011vu} as we will see below. The WIMP-nucleon cross sections $\sigma_{p,n}^{\rm SI}$ (the quantity in which experimental results are often quoted) are related to the $f_{p,n}$ by $\sigma_{p,n}^{\rm SI} = 4 \mu_{p,n}^2 f_{p,n}^2 / \pi$, where $\mu_p$ ($\mu_n$) is the reduced mass of the WIMP-proton (WIMP-neutron) system.

For SD scattering, the WIMP-nucleus cross section can be written as
\begin{equation} \label{eq:diffSD}
	\sigma_0^{\rm SD} = \frac{4}{\pi} \mu^2 \frac{J\left(J+1\right)}{J^2} \left( a_p \left\langle S_p \right\rangle + a_n \left\langle S_n \right\rangle \right)^2 \;,
\end{equation}
where $J$ is the spin of the target nucleus, $\left\langle S_p \right\rangle$ ($\left\langle S_n \right\rangle$) the average spin carried by the protons (neutrons) in the target nucleus, and $a_p$ ($a_n$) is the effective WIMP-proton (WIMP-neutron) coupling. Note, that unlike the SI case, $a_p$ and $a_n$ differ substantially in typical WIMP DM models. The $a_{p,n}$ are related to the WIMP-nucleon cross sections by $\sigma_{p,n}^{\rm SD} = 12 \mu_{p,n}^2 a_{p,n}^2/\pi$.

The differential recoil rate in Eq.~\eqref{eq:diffRate} also depends on astrophysics through the local WIMP density $\rho_\chi$ and the velocity distribution $f(\mathbf{v},t)$. For the purpose of this work, we adopt a Gaussian isotropic distribution in the galactic rest frame truncated at the galactic excape velocity $v_{\rm esc}$ as in the Standard Halo Model (SHM)~\cite{Freese:2012xd}, boosted to the Earth's frame,
\begin{equation}
	f(\mathbf{v},t) = \frac{1}{N_{\rm esc} \left(2\pi\sigma_v^2\right)^{3/2}} e^{-\frac{\left|\mathbf{v}+\mathbf{v}_E\right|^2}{2\sigma_v^2}} \Theta\left(v_{\rm esc} - \left|\mathbf{v}+\mathbf{v}_E\right|\right) \;,
\end{equation}
where $\mathbf{v}_E$ is the velocity of the Earth with respect to the galactic rest frame and $\sigma_v$ the velocity dispersion. The normalization factor is given by
\begin{equation}
	N_{\rm esc} = {\rm erf}\left(\frac{v_{\rm esc}}{\sqrt{2}\sigma_v}\right) - \sqrt{\frac{2}{\pi}} \frac{v_{\rm esc}}{\sigma_v} e^{-\frac{v_{\rm esc}^2}{2 \sigma_v^2}} \;.
\end{equation}
The velocity integral in Eq.~\eqref{eq:diffRate2}, i.e. the mean inverse speed, can then be written as~\cite{Gondolo:2002np,Catena:2015vpa} 
\begin{widetext}
\begin{equation} \label{eq:eta}
	\eta(E_{\rm nr},t) = \int\limits_{v_{\rm min}} d^3v \,\frac{f(\mathbf{v},t)}{v} = \frac{1}{N_{\rm esc} \sqrt{2\pi\sigma_v^2}} \int\limits_{-1}^1 d\left(\cos\theta\right)\,\left[ e^{-\frac{\left(v_{\rm min} + v_E \cos\theta \right)^2 }{2\sigma_v^2}} - e^{-\frac{v_{\rm esc}^2}{2\sigma_v^2}} \right] \Theta\left[ v_{\rm esc} - \left( v_{\rm min} + v_E \cos\theta \right) \right]\;.
\end{equation}
\end{widetext}
The time dependence of $\eta(E_{\rm nr},t)$, in turn gives rise to a time dependence of the differential recoil rate, due to the time dependence of the Earth's speed,
\begin{equation}
	v_E(t) = v_\odot + v_{\rm orb} \cos\gamma \cos\left[ \omega \left(t-t_0\right) \right] \;.
\end{equation}
Here the speed of the Sun with respect to the galactic rest frame, $v_\odot$, can safely be assumed to be constant on the timescale of order years relevant for direct detection experiments, and $v_{\rm orb}$ is the orbital speed of the Earth around the Sun, where $\cos\gamma \approx 0.49$ accounts for the inclination of the Earth's orbit with respect to the Sun's trajectory in the galactic rest frame. The angular frequency of the Earth's rotation is $\omega = 2\pi/{\rm year}$, and $t_0$ is the date at which $v_E$ is maximal, around June 2nd.

Note, that there are significant uncertainties on the astrophysical parameters relevant for the SHM. In the following, we adopt fiducial values of $\rho_\chi = 0.3\,$GeV/cm$^3$, $\sigma_v = 166\,$km/s, $v_{\rm esc} = 550\,$km/s, and $v_\odot = \sqrt{2}\sigma_v + 13\,{\rm km/s} = 248\,$km/s. 

In Fig.~\ref{fig:mod} we show the characteristics of the annual modulation signal for a single nuclear target in the detector. Since the time dependence of the recoil rate is due to the time dependence of the mean inverse speed $\eta(E_{\rm nr},t)$, one can parameterize the annual modulation signal in terms of the time dependence of $\eta$. In the left panel we show the difference of $\eta$ at the times when $v_E$ is maximal and when $v_E$ is minimal as a function of $v_{\rm min}$. Recall that for each set of values of the target nucleus and DM mass, $v_{\rm min}$ is directly related to the recoil energy via $v_{\rm min} = \sqrt{M E_{\rm nr}/2\mu^2}$. For large $v_{\rm min}$, the amplitude of the modulation is positive. The amplitude is growing with decreasing $v_{\rm min}$ and becomes maximal for $v_{\rm min} \approx 360\,$km/s. For even smaller $v_{\rm min}$ the amplitude is decreasing with decreasing $v_{\rm min}$, and changes sign for $v_{\rm min} \approx 210\,$km/s. 

Heuristically, this behavior can be understood from the shape of the velocity distribution: while $f(\mathbf{v},t)$ is an isotropic Gaussian distribution in the galactic rest frame with maximum at $v = 0$, once boosted into the Earth's frame it becomes a more peaked distribution with maximum at $v = v_E$, dropping quickly for larger or smaller $v$. The integral $\eta$ is dominated by contributions of the integrand $f(\mathbf{v},t)/v$ close to $v_{\rm min}$. The rotation of the Earth around the Sun corresponds to shifting $f(\mathbf{v},t)$ to larger (when $v_E$ becomes maximal) or smaller (when $v_E$ becomes minimal) values of $v$. Thus, for $v_{\rm min}$ much larger than $v_E$, the modulation amplitude is positive since more WIMPs with sufficiently large $v$ are available when $v_E$ is largest. When $v_{\rm min}$ approaches $v_E$, this effect becomes smaller because the integral $\eta$ is dominated by the values of $f(\mathbf{v},t)/v$ where $f({\mathbf{v},t})$ is maximal for both the largest and smallest values of $v_E$ during the year. Finally, when $v_{\rm min}$ becomes somewhat smaller than $v_E$, the effect reverses since $v_{\rm min}$ is now on the sharply falling side of $f(\mathbf{v},t)$: Since the integral $\eta$ is dominated by values close to $v_{\rm min}$, less WIMPs with the proper $v$ are now available when $v_E$ is largest, which leads to change of the sign of the modulation amplitude for $v_{\rm min} \lesssim 210\,$km/s.

In the right panel of Fig.~\ref{fig:mod} we show the shape of the modulated signal as a function of time for different values of $v_{\rm min}$. As long as $v_{\rm min}$ is not very close to the value for which the modulation amplitude changes sign, the modulation signal is well approximated by a single cosine. However, for $v_{\rm min}$ close to $v_{\rm min} \approx 210\,$km/s, the cosine approximation is no longer valid, see also Refs.~\cite{Chang:2011eb,Lee:2013xxa}.

This behavior of the modulation signal offers a unique test of the compatibility of a WIMP signal with an annual modulation signal observed in a direct detection experiment: For sufficiently low recoil energy threshold, one should 1) observe a decreasing modulation amplitude and finally the modulation amplitude switching sign, and 2) in the vicinity of the recoil energies corresponding to $v_{\rm min} \approx 210\,$km/s the modulation pattern should not follow a single cosine approximation. If both of these statements hold, the observed signal is consistent with a) a WIMP interacting via canonical velocity independent interactions and b) the local WIMP velocity being well approximated by the SHM. If one or both of the statements 1) and 2) are not satisfied, then one of the assumptions a) or b) must be wrong.

\section{Analysis strategy}\label{sec:Analysis}

\begin{table}
	\setlength{\tabcolsep}{8pt}
	\begin{tabular}{ccc}
		\hline\hline
		Energy & \multicolumn{2}{c}{Average $S_m$} \\
		$[ {\rm keV}_{\rm ee}]$ & \multicolumn{2}{c}{$[{\rm counts/day/kg/keV}_{\rm ee}]$} \\
		& phase1~\cite{Bernabei:2013xsa} & phase2~\cite{Bernabei:2018yyw} \\
		\hline
		$1.0 - 1.5$ & -- & $0.0232\pm0.0052$\\
		$1.5 - 2.0$ & -- & $0.0164\pm0.0043$\\
		$2.0 - 2.5$ & $0.0161\pm0.0040$ & $0.0178\pm0.0028$\\
		$2.5 - 3.0$ & $0.0260\pm0.0044$ & $0.0190\pm0.0029$\\
		$3.0 - 3.5$ & $0.0220\pm0.0044$ & $0.0178\pm0.0028$\\
		$3.5 - 4.0$ & $0.0084\pm0.0041$ & $0.0109\pm0.0025$\\
		$4.0 - 5.0$ & $0.0080\pm0.0024$ & $0.0075\pm0.0015$\\
		$5.0 - 6.0$ & $0.0065\pm0.0022$ & $0.0066\pm0.0014$\\
		$6.0 - 7.0$ & $0.0002\pm0.0021$ & $0.0012\pm0.0013$\\
		$7.0 - 20.0$ & $0.0005\pm0.0006$ & $0.0007\pm0.0004$\\
		\hline\hline
	\end{tabular}
	\caption{Average modulation amplitude $S_m$ observed by DAMA/LIBRA in the given energy bins after our rebinning of the original data from Refs.~\cite{Bernabei:2018yyw, Bernabei:2013xsa} in order to improve the signal to noise ratio in our statistical test.}
	\label{tab:rebin}
\end{table}

\begin{figure*}
	\includegraphics[width=0.49\linewidth]{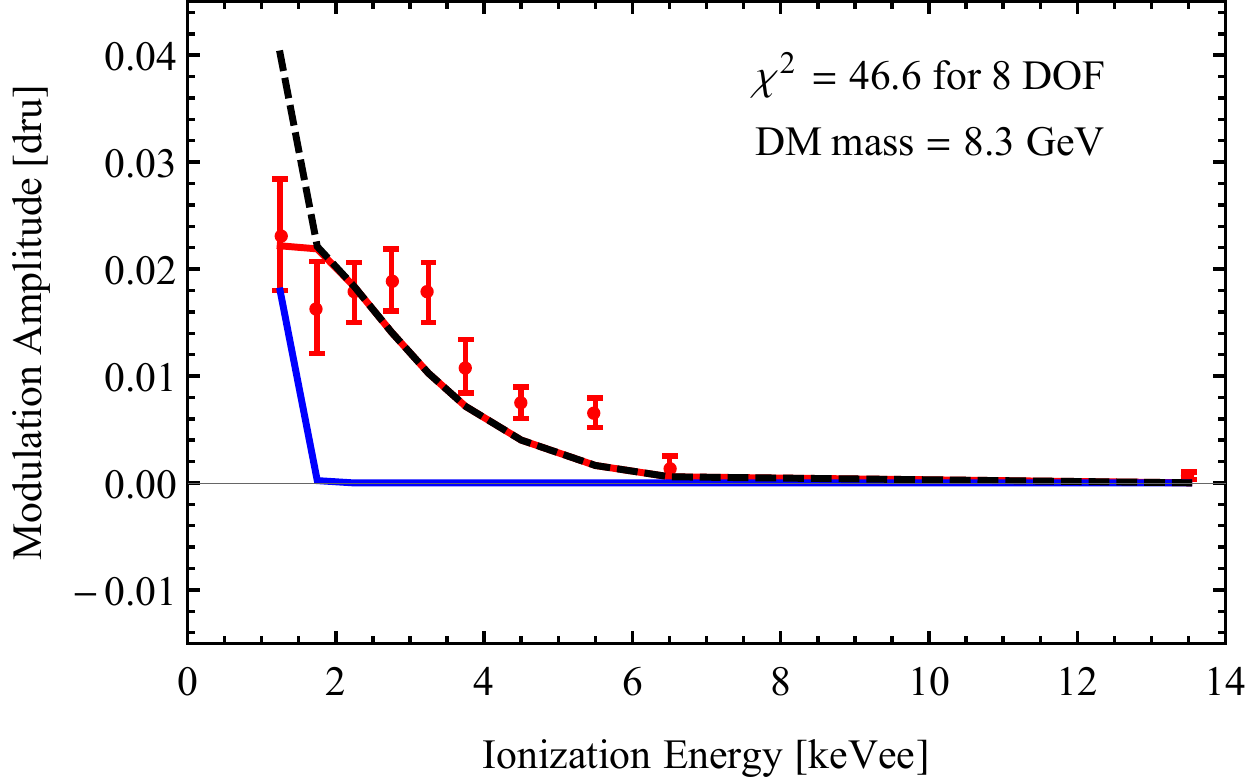}
	\includegraphics[width=0.49\linewidth]{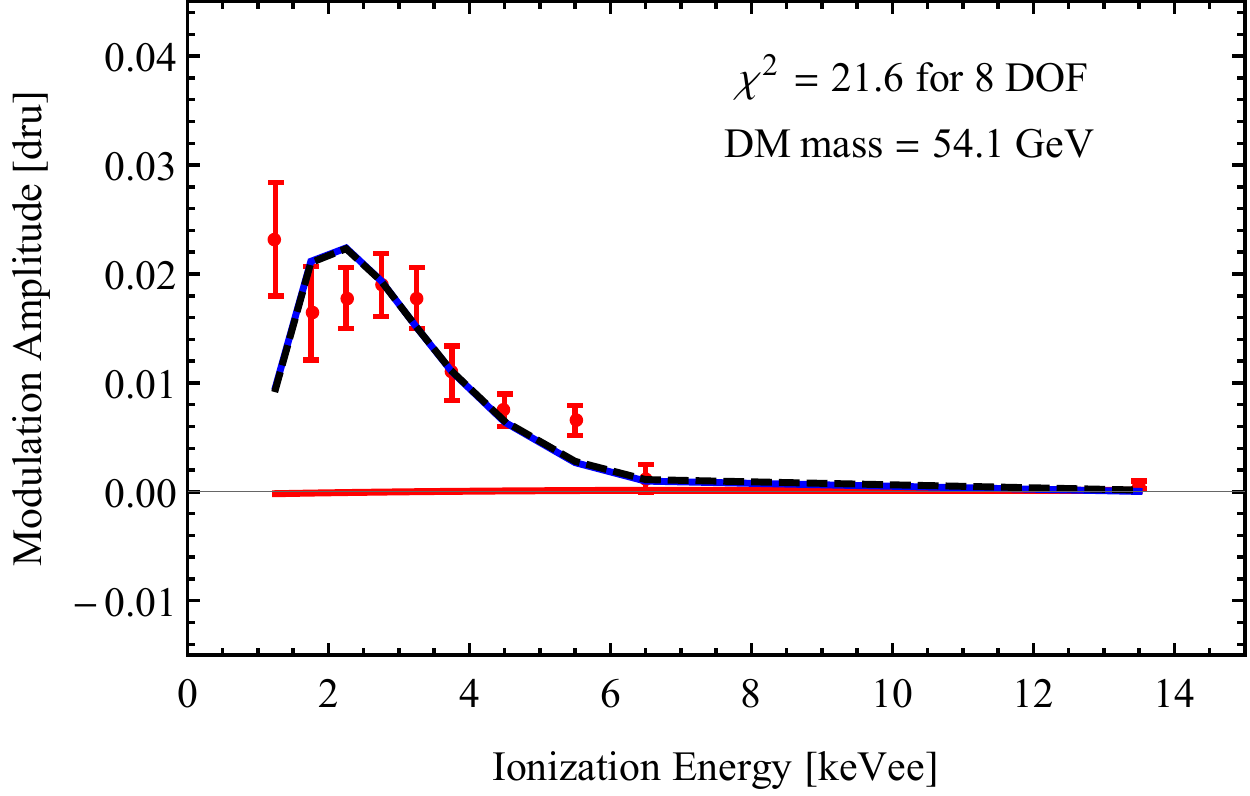}
	\caption{Modulation amplitude (in units of [counts/day/kg/keV$_{\rm ee}$]) as a function of ionization energy $E_{\rm ee}$ for the best fit points in the low mass region (left) and high mass region (right) assuming canonical (isospin conserving) SI scattering. The red data points and error bars show the DAMA/LIBRA-phase2 data after our rebinning. The red (blue) line show the contribution to the modulation amplitude from WIMPs scattering off sodium (iodine). The black dashed line shows the summed expected modulation signal. Note, that the low mass point is disfavored by $5.2\,\sigma$ and the high mass point by $2.5\,\sigma$.}
	\label{fig:SI_bestfit}
\end{figure*}

The DAMA/LIBRA collaboration present their results in terms of the count rate as function of time as well as the modulation amplitude as function of recoil energy. Both pieces of information can directly be compared to the theoretical prediction in WIMP models discussed in the previous section when taking into account the realities of an experiment. We will closely follow the analysis strategy employed in Ref.~\cite{Kelso:2013gda}. 

The DAMA/LIBRA detector does not measure the nuclear recoil energy directly, but the scintillation light produced by the nuclear recoils. The collaboration uses gamma ray sources for the calibration of their detectors, which induce electron, not nuclear, recoils. Thus, experimental results are presented by DAMA/LIBRA in terms of electron equivalent energy $E_{\rm ee}$ measured in keV$_{\rm ee}$. The nuclear recoil energy $E_{\rm nr}$ is related to $E_{\rm ee}$ by the so-called {\it quenching factor} $E_{\rm ee} = Q E_{\rm nr}$. Except where noted explicitly, we adopt the quenching factors used by the DAMA/LIBRA collaboration, $Q_{\rm Na} = 0.3$ and $Q_{\rm I} = 0.09$~\cite{Bernabei:1996vj}. Note, that there is some uncertainty on the value and energy dependence of the quenching factors, see e.g. Refs.~\cite{Collar:2013gu,Xu:2015wha,Stiegler:2017kjw}. We briefly discuss effects of energy-dependent quenching factors on our results at the end of Sec.~\ref{sec:Results}.

In addition, one has to take the finite energy resolution as well as the detector efficiency into account. We implement these effects via
\begin{equation}
	\frac{dR}{dE_{\rm ee}} (E_{\rm ee},t) = \int\limits_0^\infty d E_{\rm nr}\,\varepsilon(Q E_{\rm nr}) \phi (E_{\rm nr},E_{\rm ee}) \frac{dR}{dE_{\rm nr}}(E_{\rm nr},t) \;,
\end{equation}
where $\varepsilon(Q E_{\rm nr})$ is the detector efficiency and
\begin{equation}
	\phi(E_{\rm nr},E_{\rm ee}) = \frac{1}{2\pi\sigma^2(Q,E_{\rm nr})} e^{-\frac{\left(E_{\rm ee} - Q E_{\rm nr}\right)^2}{2 \sigma^2(Q E_{\rm nr})}}\;,
\end{equation}
is the differential response function defined such that $\phi(E_{\rm nr},E_{\rm ee}) dE_{\rm ee}$ is the probability that a nuclear recoil of energy $E_{\rm nr}$ will produce a scintillation signal measured between $E_{\rm ee}$ and $E_{\rm ee} + dE_{\rm ee}$. The energy resolution of the photo multiplier tubes at the relevant energies is given by~\cite{Bernabei:2008yh,Bernabei:2012zzb}
\begin{equation}
	\sigma(Q E_{\rm nr}) = \alpha \sqrt{Q E_{\rm nr}} + \beta Q E_{\rm nr} \;,
\end{equation}
with $\alpha = \left(0.448 \pm 0.035\right)\,\sqrt{{\rm keV}_{\rm ee}}$ and $\beta = \left(9.1\pm5.1\right)\times 10^{-3}$. Since the DAMA/LIBRA collaboration present results corrected for the efficiency, we will use $\varepsilon(Q E_{\rm nr}) = 1$ in the following.

With those corrections, we can compare theoretical predictions in WIMP models to the measured values using a $\chi^2$ test. For example, for the modulation amplitude, which is presented as binned measurements in energy bins, this takes the form
\begin{equation} \label{eq:chi2}
	\chi^2(m_\chi, \sigma_0) = \sum\limits_k \frac{\left[ S_{m,k} - S_{m,k}^T(m_\chi,\sigma_0) \right]^2}{\sigma_k^2} \;,
\end{equation}
where $S_{m,k}$ ($S_{m,k}^T$) is the measured (theoretically expected) modulation amplitude averaged over the bin $k$, $\sigma_k$ is the measurement uncertainty, and $m_\chi$ and $\sigma_0$ are the WIMP mass and WIMP-nucleus scattering cross section discussed in Sec.~\ref{sec:AnnMod}.

The {\it goodness of fit} of the WIMP hypothesis is tested by minimizing the $\chi^2$ over the WIMP parameter space, comparing the resulting $\chi_{\rm min}^2$ at the local minima with the number of degrees of freedom ($dof$), given by the number of data bins minus the number of fitted parameters, and ensuring that the $\chi^2$ in Eq.~\eqref{eq:chi2} follows a $\chi^2$ distribution with the appropriate $dof$ in the vicinity of the local minimum. In addition, confidence regions are determined using the goodness of fit test. 

In order to increase the signal to noise ratio, we rebin the DAMA/LIBRA data into 10 signal bins, as given in Tab.~\ref{tab:rebin}. This choice of the bins is motivated by two reasons: 1) at larger recoil energies, the width of the bins chosen by the DAMA/LIBRA collaboration is much smaller than the energy resolution of the detector, and 2) the WIMP signal at large recoil energies is negligible compared to the signal at small recoil energies. See Ref.~\cite{Kelso:2013gda} for a detailed discussion and motivation of this binning scheme.

\section{Results}\label{sec:Results}

We test the compatibility of three WIMP DM hypotheses with the modulation signal observed by the DAMA/LIBRA collaboration: 1) canonical SI scattering where $f_p = f_n$, 2) isospin violating (IV) scattering (which is also independent of spin) where we allow for isospin violation $f_p \neq f_n$, and 3) SD scattering. In case 1), we fit two model parameters: the WIMP mass $m_\chi$ and the WIMP-proton scattering cross section $\sigma_p^{\rm SI}$. In case 2), we are fitting three model parameters: the WIMP mass $m_\chi$, the WIMP-proton cross sections $\sigma_p^{\rm SI}$, and the ratio of the effective couplings to neutrons and protons $f_n/f_p$. In case 3) we fit either two model parameters, $m_\chi$ and $\sigma_p^{\rm SD}$ ($\sigma_n^{\rm SD}$), when assuming proton-only (neutron-only) SD couplings or the three parameters $m_\chi$, $\sigma_p^{\rm SD}$ and the ratio of the effective couplings to neutrons and protons $a_n/a_p$ when allowing for the general mixed SD case. 

{\it Regarding canonical SI scattering (with no isospin violation)}:
With the older DAMA/LIBRA data, canonical isospin conserving SI scattering allowed for a good fit in both a low ($\sim10\,$GeV) and high mass ($\sim 70\,$GeV) range for WIMPs.  With the new data, the best fit high mass shifts to $\sim 54$ GeV and the best fit low mass to $\sim 8\,$GeV, cf. Fig.~\ref{fig:SI_bestfit}. However, the new data do not allow for a good fit to the observed modulation signal. In the new best fit regions, SI scattering is disfavored at $5.2\,\sigma$ for the low mass region around $\sim 8$ GeV and at $2.5\,\sigma$ for the high mass region around $\sim 54\,$GeV. The new data play important roles in these conclusions for the following two reasons: the phase2 results contains two new bins for recoil energies between $1.0$ and $2.0$\,keV$_{\rm ee}$, and, the modulation signal in the two bins between $2.5$ and $3.5\,$keV$_{\rm ee}$ has shifted to lower values, cf. Tab.~\ref{tab:rebin}. While the previous phase1 data seemed to hint at a decreasing modulation amplitude below $\sim 2.5\,$keV$_{\rm ee}$, the updated results with the inclusion of the phase2 data indicate a modulation amplitude smoothly rising with smaller recoil energies. As discussed in Ref.~\cite{Kelso:2013gda}, this behavior disfavors isospin conserving SI scattering for both the previous low and high mass regions due to a conspiracy of different effects: For the low mass hypothesis, the modulation amplitude is dominated by scattering off sodium for recoil energies $\gtrsim 1.5\,$keV$_{\rm ee}$. However, for a 10\,GeV WIMP, the modulation amplitude from scattering with sodium has its maximum at $E_{\rm ee} \approx 2\,$keV$_{\rm ee}$, and decreases for lower recoil energies, approaching a phase reversal. 
At roughly the recoil energy of the sodium phase reversal and below, scattering off iodine becomes relevant, and the corresponding modulation amplitude rises sharply for smaller recoil energies. Summing both effects would lead to a sharply rising modulation amplitude in the lowest bin of the phase2 data.

\begin{figure}
	\includegraphics[width=\linewidth]{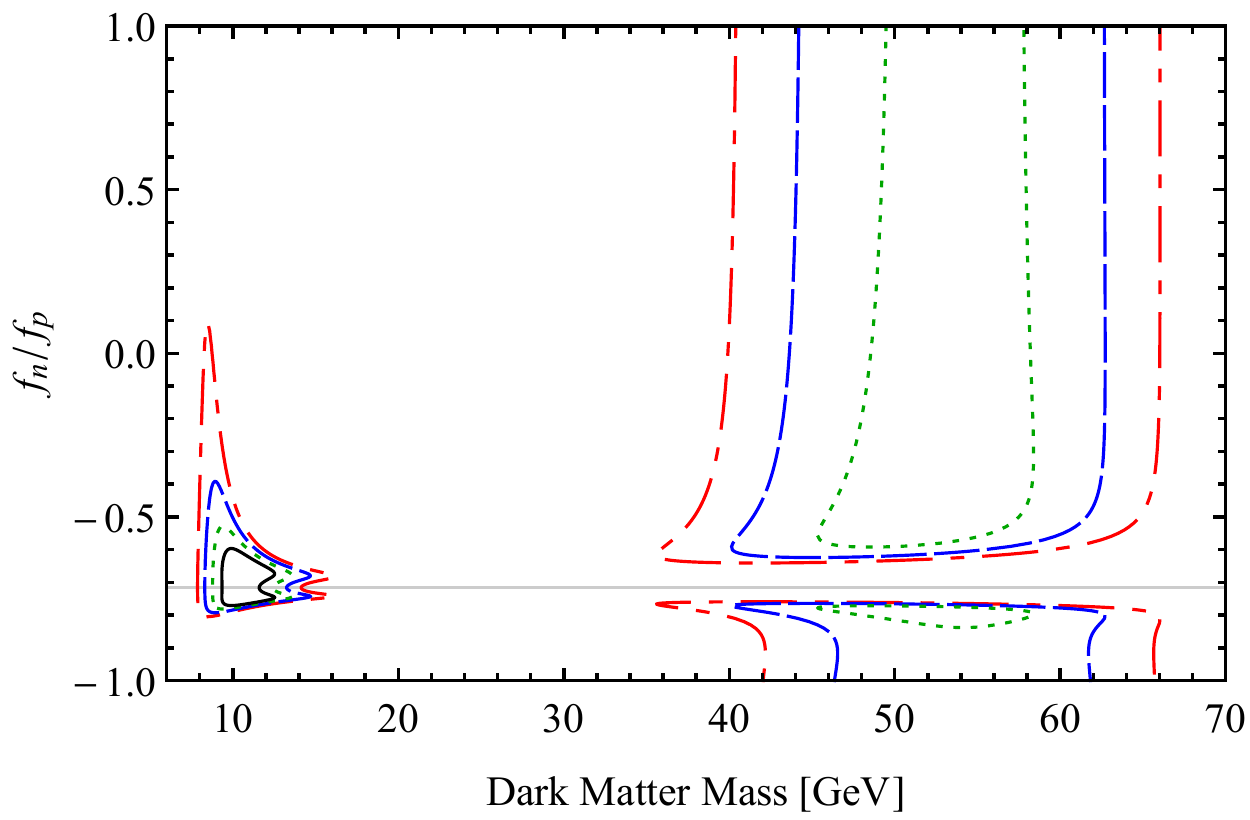}
	\caption{Goodness of fit regions for the case of isospin violating (IV) couplings (still spin-independent but with $f_n/f_p \neq 1$). The solid black, dotted green, dashed blue, and dash-dotted red contours correspond to the 2, 3, 4, and 5\,$\sigma$ regions, respectively. The best fit point in the low mass region has a DM mass of $m_\chi = 10.7\,$GeV, $f_n/f_p = -0.666$, a WIMP-proton cross section of $\sigma_p^{\rm SI} = 1.05 \times 10^{-2}\,$pb, and a reduced $\chi^2$ value of $\chi^2_r = 7.32/7$. The best fit point in the high mass region is $m_\chi = 52.1\,$GeV, $f_n/f_p = -0.613$, $\sigma_p^{\rm SI} = 2.25 \times 10^{-3}\,$pb, and $\chi_r = 17.26/7$. The thin gray line indicates $f_n/f_p = 53/74$, the value for which the scattering cross section off iodine vanishes.}
	\label{fig:IVDM_results}
\end{figure}

For the high mass hypothesis, the modulation signal is dominated by scattering off iodine in the entire energy range relevant for the observed signal. For a 50\,GeV WIMP scattering off iodine, the modulation amplitude has its maximum at $E_{\rm ee} \approx 2.7\,$keV$_{\rm ee}$, hence, this would lead to a decreasing modulation amplitude in the lowest bins of the phase2 data. 

However, the phase2 data do not display either of the behaviors described above, but instead indicate a modulation amplitude smoothly rising with smaller recoil energies.

{\it Regarding Isospin Violating (IV) Dark Matter:}
In order to fit the data, the scattering amplitude off iodine must be suppressed with respect to the scattering rate off sodium, which is possible when allowing for isospin violation. Since sodium and iodine are almost exclusively composed of the isotopes $^{23}$Na and $^{127}$I, respectively, the scattering cross section off iodine is suppressed if the ratio of the effective WIMP-neutron and WIMP-proton couplings is approximately $f_n/f_p=-53/74 \approx -0.716$, for which the effective coupling to $^{127}$I vanishes. Varying the value from the isospin conserving case $f_n/f_p = 1$ to $f_n/f_p = -53/74$ alters the relative strength of the decreasing modulation amplitude from scattering off sodium and the increasing amplitude from scattering off iodine at low recoil energies in the low mass case, such that a much better fit to the observed signal is obtained. Similarly, for the high mass case suppressing the contribution from scattering off iodine allows for a better fit to the data because one finds a non-negligible contribution to the modulation signal from scattering off sodium for the largest energy bins in the DAMA data.
 
Fig.~\ref{fig:IVDM_results} shows our results for spin-independent scattering which allows for isospin violation. We find two minima in the $\chi^2$ distribution: a {\it low mass} region with a best fit point with $m_\chi = 10.7\,$GeV, $f_n/f_p = -0.666$, a WIMP-proton cross section of $\sigma_p^{\rm SI} = 1.05 \times 10^{-2}\,$pb, and a reduced $\chi^2$ value of $\chi^2_r = 7.32/7$, and a {\it high mass} region with a best fit point $m_\chi = 52.1\,$GeV, $f_n/f_p = -0.613$, $\sigma_p^{\rm SI} = 2.25 \times 10^{-3}\,$pb, and $\chi_r = 17.26/7$. The reduced $\chi^2$ values correspond to $0.85\,\sigma$ for the low-mass best fit point and $2.4\,\sigma$ for the high-mass best fit point. Both best fit regions differ significantly from the best fit regions for the previous DAMA/LIBRA-phase1 data in the canonical SI case derived with a similar binning scheme~\cite{Kelso:2013gda}.  

\begin{figure}
	\includegraphics[width=\linewidth]{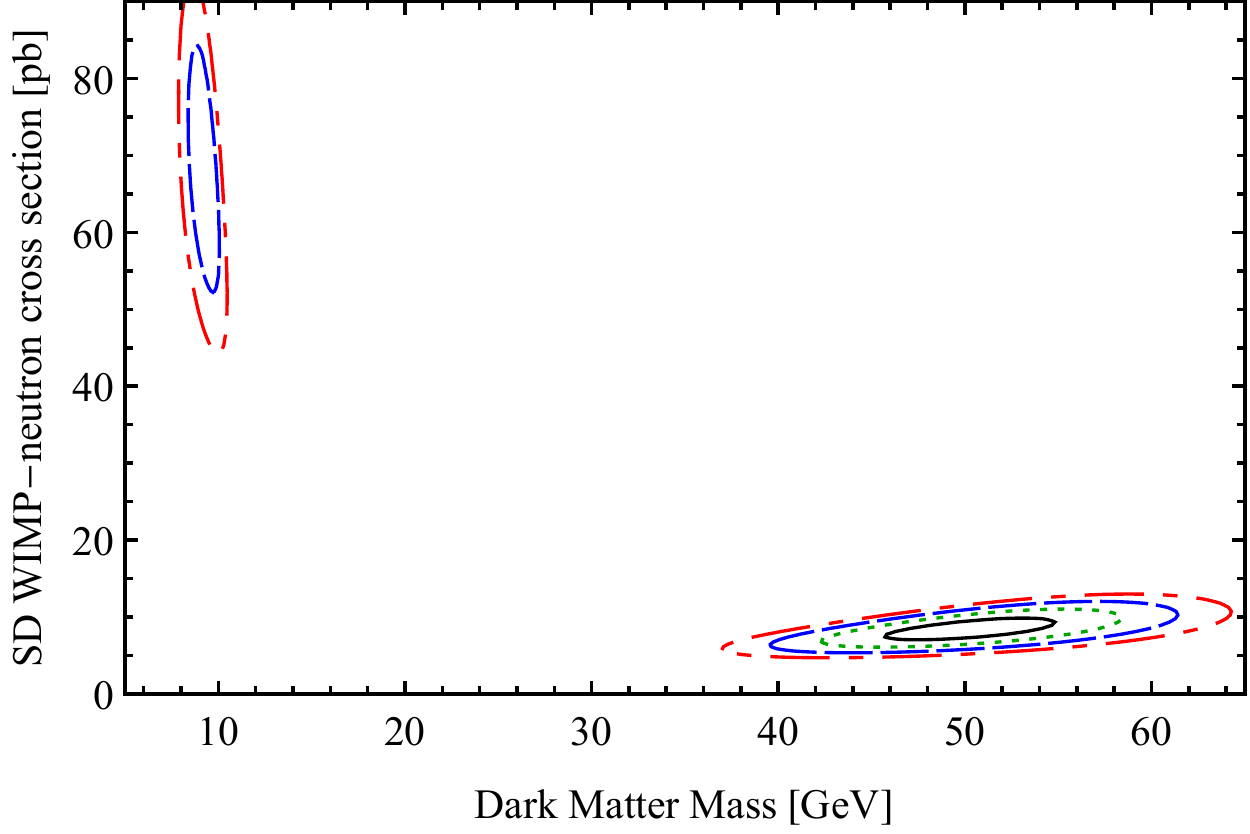}
	\includegraphics[width=\linewidth]{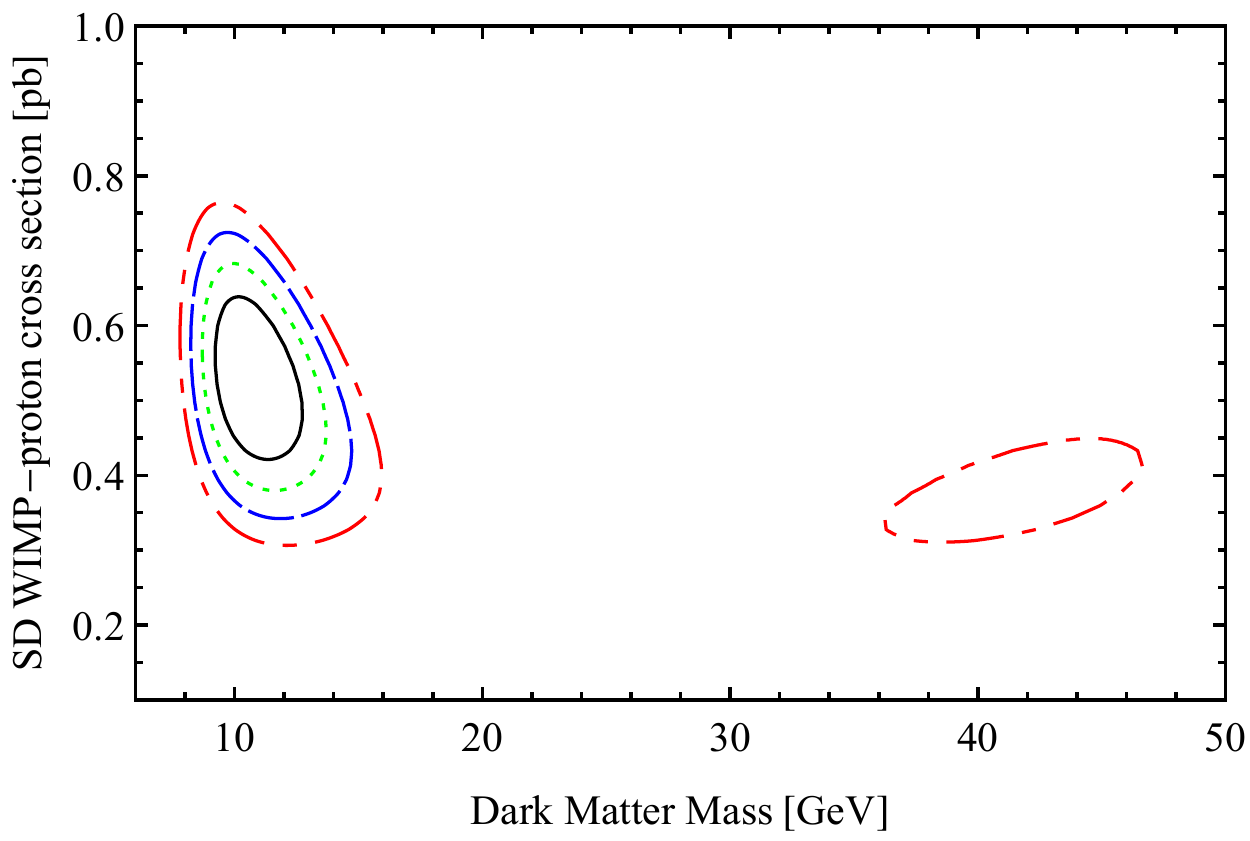}
	\caption{Goodness of fit regions for SD-neutron (top) and SD-proton (bottom) couplings. The different lines refer to the same confidence levels as in Fig.~\ref{fig:IVDM_results}. The best fit points are listed in Tab.~\ref{tab:SDbestfit}.}
	\label{fig:SD_np_only}
\end{figure}

{\it Regarding SD scattering:}
For the SI (IV) case, the ratio of the contributions from sodium and iodine to the event rate in DAMA is given by the WIMP mass and the number of nucleons (number of protons and neutrons together with the ratio $f_n/f_p$). For SD interactions, the relative contributions from sodium and iodine are instead controlled by the WIMP mass, the fraction of the nuclear spin carried by protons and neutrons, respectively, and the ratio between the WIMP-neutron and WIMP-proton SD couplings $a_p/a_n$. For both sodium and iodine most of the nuclear spin is carried by the protons. Note that there are significant theoretical uncertainties in the computations of the form factors describing which fraction of the nuclear spin is carried by protons or neutrons, respectively, in particular for heavy nuclei such as iodine. Shifts in the values of the form factors would have minor quantitative effects on our result but still allow for acceptable fits. 

\begin{figure}
	\includegraphics[width=\linewidth]{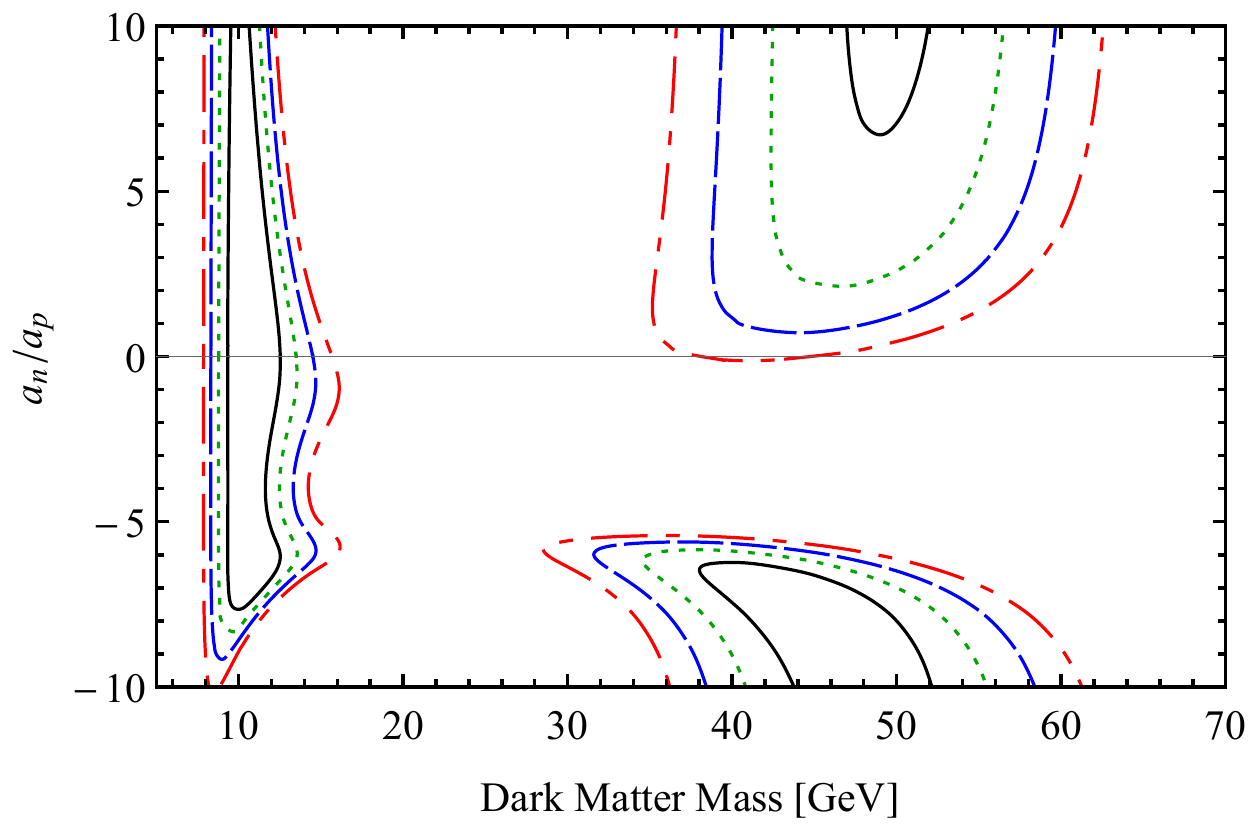}
	\caption{Same as Fig.~\ref{fig:IVDM_results}, but for the mixed SD case. The best fit points are listed in Tab.~\ref{tab:SDbestfit}.}
	\label{fig:SD_results}
\end{figure}

\begin{table}
	\begin{tabular}{ccccc}
		\hline\hline
		& $m_\chi$ [GeV] & $\sigma_i^{\rm SD}$ [pb] & $a_n/a_p$ & $\chi^2/dof$ \\
		\hline
		\multirow{2}{*}{$n$-only} & 9.3 & 68.0 & $\infty$ & 24.5/8 ($3.1\,\sigma$) \\
			& 50.2 & 8.4 & $\infty$ & 12.0/8 ($1.4\,\sigma$) \\
		\hline
		\multirow{2}{*}{$p$-only} & 10.8 & 0.53 & 0 & 7.4/8 ($0.7\,\sigma$) \\
			& 41.5 & 0.38 & 0 & 39.8/8 ($4.6\,\sigma$) \\
		\hline
		\multirow{3}{*}{mixed} & 10.7 & 1.9 & -6.32  & 7.3/8 ($0.7\,\sigma$) \\
			& 10.8 & 0.47 & 0.665  & 7.3/8 ($0.7\,\sigma$) \\
& 44.2 & 0.43 & -7.46  & 9.6/8 ($1.0\,\sigma$) \\

		\hline\hline
	\end{tabular}
	\caption{Best fit points for the different SD hypotheses. The first two best-fit points are for neutron-only couplings where $\sigma_i^{\rm SD} = \sigma_p^{\rm SD}$ refers to the WIMP-neutron SD coupling. The third and fourth line are for the proton only case where $\sigma_i^{\rm SD} = \sigma_n^{\rm SD}$. The last three entries are for the mixed case where in addition to the reference cross section and the Dark Matter mass $m_\chi$ we fit the ratio of the effective couplings to neutrons and protons $a_n/a_p$. Here, $\sigma_i^{\rm SD} = \sigma_p^{\rm SD}$.}
	\label{tab:SDbestfit}
\end{table}

We find best-fit points with acceptable reduced $\chi^2$ values in all cases we consider: proton-only, neutron-only, and mixed SD couplings, cf. Tab.~\ref{tab:SDbestfit}. In all SD cases considered we find a low mass best fit region around $10\,$GeV and a high mass region around $\sim 45\,$GeV. For neutron-only (proton-only) couplings the high mass (low mass) region is preferred, while for the mixed case both regions yield a good fit to the data. 

If Fig.~\ref{fig:SD_np_only} we show the goodness of fit regions for the neutron-only and  proton-only case. Comparing these two cases we find that the best fit regions correspond to WIMP-neutron cross sections approximately two orders of magnitude larger in the neutron-only case than the WIMP-proton cross sections in the proton-only case. Since the protons carry most of the nuclear spin for both sodium and iodine, such cross sections yield comparable scattering rates and modulation signals for proton-only and neutron-only couplings, cf. Eq.~\eqref{eq:diffSD}. We can also note that for the neutron-only case the high mass region yields a better fit than the low mass region, while the situation is opposite in the proton-only case. This is again due to the spin carried by the neutrons and protons, respectively, and the required balance between scattering off sodium and iodine in order to yield an energy dependence of the signal matching the DAMA/LIBRA data: The ratio of the spin carried by the neutrons at zero momentum for sodium and iodine is $\langle S_n^{\rm Na}\rangle/\langle S_n^{\rm I}\rangle \sim 0.3$, while the ratio of the spin carried by the protons at zero momentum is given by $\langle S_p^{\rm Na}\rangle/\langle S_p^{\rm I}\rangle \sim 0.7$~\cite{Bednyakov:2004xq}. The contribution from scattering off iodine to the total modulation signal must be suppressed more strongly with respect to the contribution from sodium in the low mass region than in the high mass region; recall the discussion of IV couplings and Figs.~\ref{fig:SI_bestfit} and~\ref{fig:IVDM_results}.

Fig.~\ref{fig:SD_results} shows the goodness of fit regions for the mixed SD case. The neutron-only (proton-only) case corresponds to the limit $|a_n/a_p|\to\infty$ ($a_n/a_p\to 0$). Much of the shape of the best fit regions can be understood from comparing to the best fit regions for the proton-only and neutron-only cases (cf. Fig.~\ref{fig:SD_np_only}) and the form factor for iodine.  For $a_n/ap\lesssim -5$, the form factor for iodine produces a much longer tail to the dark matter signal.  As the data exhibit a slow rise, this leads to the best fits.  

For the low mass region, the situation is slightly more complex.  Proton only couplings yield a good fit to the data while neutron-only couplings are disfavored by $\sim 3\,\sigma$. Thus, in the mixed case, small values of $a_n/a_p$ also produce a good fit.  Additionally, the closing of the contours at large negative values of $a_n/a_p$ can be understood as follows.  For moderately negative values for $a_n/a_p$ the primary scattering is from sodium.  Most of the iodine events are below the threshold, but the long tail of the iodine signal will help to add events at low energies which improves the fit to the data.  As $a_n/a_p$ is decreased even further though, too many iodine scatters are created above threshold, producing the poor fit in the low energy region.    

A word of caution is in order when directly comparing the reduced $\chi^2$ values between different coupling hypotheses. The reduced $\chi^2$ values do not account for the likelihood of the different scenarios; such factors can only properly be taken into account in the framework of specific models which allow for the different coupling scenarios to occur. Hence, such comparisons should not be taken at face values but only as qualitative statements.

Finally, we note that our main conclusions are not affected by different choices for the quenching factors. In particular, we repeated our analysis using the energy-dependent quenching factor for sodium and a constant quenching factor $Q_I = 0.04$ for iodine as found in Ref.~\cite{Collar:2013gu}. In the case of canonical SI interactions such values of the quenching factors make the fit to the new DAMA data more difficult. Recall that the difficulty in fitting the DAMA/LIBRA-phase2 data stems from the contribution of iodine recoils becoming too large for small recoil energies. This problem is worsened by the energy dependence of the sodium quenching factor found in Refs.~\cite{Collar:2013gu,Xu:2015wha,Stiegler:2017kjw}, because their results indicate that the quenching factor for sodium decreases with decreasing recoil energy. Numerically, we find that with the quenching factors from Ref.~\cite{Collar:2013gu}, canonical (isospin-conserving) SI interactions are disfavored at $7.6\,\sigma$. For isospin-violating (IV) interactions, there is only one region yielding a good fit to data, namely for DM masses $\sim 21\,$GeV and a ratio of the effective WIMP-neutron and WIMP-proton couplings close to the value where the effective WIMP-iodine coupling vanishes, $f_n/f_p \approx -0.72$. We find similar behavior for SD scattering. In the neutron-only case, we do not find a good fit in the low-mass region anymore. For heavier WIMPs we find a good fit for $m_\chi \sim 90\,$GeV, instead of $m_\chi \sim 50\,$GeV with the quenching factors used in the main analysis. For the proton-only case, the best fit-point for the low-mass region is at $m_\chi \sim 20\,$GeV with the quenching factors from Ref.~\cite{Collar:2013gu}, instead of $m_\chi \sim 10\,$GeV as found in Fig.~\ref{fig:SD_np_only}. In the mixed SD case, we find a good fit to the data for $m_\chi \sim 20\,$GeV and $a_n/a_p \sim -4$. In the high mass regions we find a good fit for masses approximately double the high mass region in Fig.~\ref{fig:SD_results} but similar values of $a_n/a_p$.

\section{Conclusions}\label{sec:Conclusions}
In this work, we have discussed the compatibility of the recently published DAMA/LIBRA-phase2 results with the Dark Matter hypothesis. We find, that the observed annual modulation signal is no longer well fitted by canonical (isospin conserving) spin-independent WIMP nucleon couplings in the SHM, as has also been noted in the revised version of Ref.~\cite{Kahlhoefer:2018knc}. For the low mass hypothesis of a $\sim 10\,$GeV WIMP which previously was a good fit to the DAMA/LIBRA-phase1 data, one would expect a sharp rise of the modulation amplitude for recoil energies below $\sim 2\,$keV$_{\rm ee}$ (for such recoil energies scattering off iodine becomes relevant while for larger recoil energies the modulation signal would predominantly be due to scattering off sodium). For the previous high mass hypothesis of a $\sim 70\,$GeV WIMP the modulation signal is predominantly due to scattering off iodine for the relevant range of recoil energies observed by DAMA/LIBRA, and one expects a decreasing modulation amplitude below $\sim 2.5\,$keV$_{\rm ee}$ recoil energies because the modulation signal approaches the change in the sign of the modulation amplitude. 
Instead of either of these behaviors, the observed modulation signal seems to be smoothly rising with decreasing recoil energies when including the new DAMA/LIBRA-phase2 data.  

The value of the best fit mass for the canonical SI case has shifted due to the new lower threshold data, yet neither the low nor high mass region remains a good fit. The SI case is disfavored by $5.2\,\sigma$ for  DM masses of $\sim8\,$GeV and $2.5\,\sigma$ for DM masses of $\sim 54\,$GeV, the new best fit masses for the SI case given the new data. 

When we consider isospin violating (IV) spin independent scattering, we find a good fit to the data for the low mass region $\sim 10\,$GeV. The best fit point requires significant isospin violation of the WIMP-nucleon coupling yielding suppressed effective WIMP-iodine couplings, namely, ratios of the WIMP-neutron to WIMP-proton couplings of $f_n/f_p = -0.666$. This value is close to $f_n/f_p = -0.716$, for which the effective WIMP-iodine coupling would essentially vanish. 

In the case of SD couplings we find regions yielding a good fit to the data in all cases considered: neutron-only, proton only, and the general mixed coupling case. In all cases, we find a $\sim 10\,$GeV low mass and a $\sim 45\,$GeV high mass region.

The difficulty in fitting the DAMA/LIBRA-phase2 data is exacerbated for energy-dependent quenching factors for sodium as found in Refs.~\cite{Collar:2013gu,Xu:2015wha,Stiegler:2017kjw}. For the quenching factors found in Ref.~\cite{Collar:2013gu} the canonical SI case is disfavored at $7.6\,\sigma$. For IV couplings as well as for all SD cases considered here we find regions with good fit to the data using the quenching factors from Ref.~\cite{Collar:2013gu}.

We have also pointed out how in the case of a observation of decreasing modulation amplitude, the temporal form of the modulation signal for recoil energies close to the energy for which the amplitude switches sign may be used to test the compatibility of the signal with the WIMP hypothesis. We predict a deviation from the simple cosine approximation for the modulated signal for such recoil energies even for the SHM. 

Finally, let us note that currently many direct detection experiments using different target nuclei rule out these models by many orders of magnitude. Additionally, multiple efforts are underway to independently test the DAMA/LIBRA results by searching for an annual modulation signal using NaI crystals~\cite{Shields:2015wka,Amare:2014jta,Fushimi:2015sew,Angloher:2016ooq,Adhikari:2017esn}. These experiments will confirm or rule out the DAMA/LIBRA signal in the coming years. If they confirm the observed modulation signal, the combination of these experiments with DAMA/LIBRA will allow for a powerful test of the WIMP Dark Matter hypothesis.

\section*{Note added}
Since the submission of the first version of this paper, Ref.~\cite{Bernabei:2018yyw} appeared. For this version, we updated our numerical results using data taken from Ref.~\cite{Bernabei:2018yyw} instead of the preliminary DAMA/LIBRA-phase2 results taken from Ref.~\cite{DAMAp2} used in the original version of our work. Furthermore, while the first version of this paper included results only for SI and IV WIMP-nucleus scattering, this version includes additional results for SD WIMP-nucleus interactions.

\begin{acknowledgments}
 SB and KF acknowledge support by the Vetenskapsr\r{a}det (Swedish Research Council) through contract No. 638-2013-8993 and the Oskar Klein Centre for Cosmoparticle Physics.
KF acknowledges support from DoE grant DE-SC007859 and the LCTP at the University of Michigan. 
SB would like to thank the  LCTP and the University of Michigan, where this work was carried out, for hospitality.
The work of CK is supported in part by Space Florida and the National Aeronautics and Space Administration through the University of Central Florida's NASA Florida Space Grant Consortium.
\end{acknowledgments}

\bibliography{theBib}

\end{document}